# Nearest-Neighboring Pairing of Monolayer NbSe$_2$ Facilitates the Emergence of Topological Superconducting States

Y. Z. Li[a], Q. Gao[a], Y. R. Li[a,b], J. X. Zhong[a,b,c]✉ and L. J. Meng[a,b]✉

[a]School of Physics and Optoelectronics, Xiangtan University, Xiangtan 411105, Hunan, People's Republic of China

[b]Hunan Key Laboratory for Micro-Nano Energy Materials and Devices, Hunan, People's Republic of China

[c]College of Sciences, Shanghai University, Shanghai 200444, People's Republic of China

**Abstract**

NbSe$_2$, which simultaneously exhibits superconductivity and spin-orbit coupling, is anticipated to pave the way for topological superconductivity and unconventional electron pairing. In this paper, we systematically study topological superconducting (TSC) phases in monolayer NbSe$_2$ through mixing on-site s-wave pairing (p$_s$) with nearest-neighbor pairing (p$_{sA1}$) based on a tight-binding model. We observe rich phases with both fixed and sensitive Chern numbers (CNs) depending on the chemical potential (μ) and out-of-plane magnetic field (V$_z$). As the p$_{sA1}$ increases, the TSC phase manifests matching and mismatching features according to whether there is a bulk-boundary correspondence (BBC). Strikingly, the introduction of mixed wave pairing significantly reduces the critical V$_z$ to form TSC phases compared with the pure s-wave paring. Moreover, the TSC phase can be modulated even at V$_z$=0 under appropriate μ and p$_{sA1}$, which is identified by the robust topological edge states (TESs) of ribbons. Additionally, the mixed pairing influences the hybridization of bulk and edge states, resulting in a matching/mismatching BBC with localized/oscillating TESs on the ribbon. Our finding is helpful for the realization of TSC states in experiment, as well as designing and regulating TSC materials.

## 1. Introduction

Topological superconductors are extensively studied with the aim of inducing and manipulating Majorana zero modes, which is crucial for realizing topological quantum information due to non-Abelian statistical property[1-7]. Recently discovered transition metal dichalcogenides are excellent candidates to host topological superconductivity. For example, Ising superconductor 1T$_d$-PdSe$_2$ is detected large in-plane critical field more than 7 times that of the Pauli limit[8], heavily gated MoS$_2$, supporting the exotic spin-singlet p+ip-wave pairing, is a topological superconducting phase that breaks time-reversal symmetry spontaneously and possesses nonzero Chern numbers (CNs)[9], 2M-WS$_2$ presents a transition temperature T$_c$ = 8.8 K and intrinsic superconductivity[10], 2H-NbSe$_2$ takes the lead as an intrinsic superconductor that exhibits superconductivity and charge-density wave phase from bulk to monolayers and withstands exceptionally high magnetic fields far beyond the Pauli limit for superconductivity[11, 12-14], and so forth.

---

✉ The corresponding author Email: jxzhong@xtu.edu.cn
✉ The corresponding author Email: ljmeng@xtu.edu.cn

Two-dimensional TSCs with nontrivial properties exhibit topological edge states (TESs) along their edges, as dictated by the well-established bulk-boundary correspondence (BBC) principle, which associates the topological invariant with the number of TESs[2]. However, recent research progress on topological materials has revealed the sensitivity and richness of the Chern number (CN) phase diagram to variations in the chemical potential (μ), magnetic field, and superconducting order parameter amplitude.Furthermore, there is evidence of a mismatch between the CN and the number of TESs, indicating a deviation from the BBC principle[15-25]. For example, graphene, 2H$_b$-stacked bilayer transition metal dichalcogenides and bilayer bismuth lattice with on-site s-wave pairing show the TSC phases with rich and high CNs, these phases do not strictly always contain the same number of TESs as the CNs[16, 17, 21]. 2D square lattice with tetragonal D$_{4h}$ symmetry considering a mixture of on-site and off-site singlet pairing manifests non-trivial high CN, massive edge states, and zero energy modes out of high symmetry points, and the number of zero-energy modes is higher than the CN in certain cases[22]. A checkerboard-lattice model combining the Chern insulator and chiral p-wave superconductivity produces TSC with non-zero CNs, and the results clearly reveal the mismatch between the CNs and TESs[20]. Previous studies indicate that, in certain cases, the CNs and TESs of two-dimensional TSC do not always agreement with the BBC principle. However, the underlying mechanism behind the mismatchness of the CN and the number of TESs requires comprehensively investigation.

In our work, we use projection operator approach[26] for 2H-NbSe$_2$ with C$_{6v}$ point group symmetry to obtain nearest-neighbor (NN) pairing function of irreducible representation A$_1$. We investigate systematically TSC phases considering mixed pairing of s-wave $\Delta_s$ and nearest-neighbor off-site $\Delta_{sA1}$ on bulk NbSe$_2$. This exploration is conducted by applying external out-of-plane magnetic field (V$_z$) and Rashba spin-orbital coupling (RSOC) based on a Bogoliubov-de Gennes (BdG) Hamiltonian. To study topological properties of TSC phase, we initially employ efficient method[27] to compute CN as a function of μ and V$_z$. The V$_z$ lies in [0.0, 0.1]eV which is accessible by external magnetic field in experiment or constructing heterojunction with two-dimensional ferromagnet (CrX$_3$, X=Cl, Br, I)[28]. Subsequently, we calculate the TESs of zigzag and armchair ribbons to confirm TSC states of CNs phase diagram. Additionally, we investigate the effect of the NN pairing potential on the bulk state by calculating the bulk band-gap between conduction band and valance band. Finally, we present the possibility distribution ($|\psi|^2$) in real space to reveal the distribution of TESs.

## 2. Methods and model

The d-orbitals of the Nb atoms in monolayer NbSe$_2$ dominating the energy bands near the E$_F$ suggests that the superconductivity is primarily contributed by the Nb-4dbands[29]. Therefore the Hamiltonian of monolayer NbSe$_2$ must meet the symmetry requirements of the triangular sublattice of Nb atoms[11, 29, 30]. Fig. 1(a) shows the trigonal lattice of Nb atoms and the zigzag/armchair ribbons width $N_z$ and $N_a$ in monolayer NbSe$_2$. The BdG Hamiltonian of sublattice Nb, incorporating $V_z$, RSOC, s-wave and nearest-neighbor pairing in the Nambu basis $\psi=(c_{k\uparrow},c_{k\downarrow},c^\dagger_{-k\uparrow},c^\dagger_{-k\downarrow})^T$ is defined as[28]

$$H(k) = \begin{bmatrix} E_t\sigma_0 + E_{R_x}\sigma_x + E_{R_y}\sigma_y + V_z\sigma_z & i(\Delta_s + \Delta_{sA1})\sigma_y \\ -i(\Delta_s + \Delta_{sA1})^*\sigma_y & -E_t\sigma_0 - E_{R_x}\sigma_x - E_{R_y}\sigma_y - V_z\sigma_z \end{bmatrix} \quad (1)$$

where $E_t$ is the hopping term and here we consider third-neighboring form for the normal band structure. This form of $E_t$ includes the first-/second-/third-neighboring parameters $t_1 = -0.04$eV, $t_2 = -0.132$eV, $t_3 = -0.012$eV. The RSOC term is given by $E_{R_x}$ and $E_{R_y}$ with strength parameter $\alpha_R=0.1$eV throughout the paper. $V_z$ represents external out-of-plane magnetic field. $\Delta_s$ describes the s-wave superconducting pairing and $\Delta_{sA1}$ is nearest-neighbor spin-singlet basis function for the irreducible representation A$_1$ with parameter $s_{A1}$. The form of $\Delta_{sA1}$ can be obtained through initial $\phi^{start}$ (Fig. 1(a)) based on projection operator method and given as

$$\Delta_{sA1}(\mathbf{k}) = s_{A1}\left[2\cos(k_x) + 2\cos\left(\frac{1}{2}k_x + \frac{\sqrt{3}}{2}k_y\right) + 2\cos\left(\frac{1}{2}k_x - \frac{\sqrt{3}}{2}k_y\right)\right] \quad (2)$$

The bulk 2H-NbSe$_2$ realizes an s-wave topological superconductor with CNs of 3/−2/−1, and its ribbon exhibits the same number topological response in terms of TESs[28]. Fig. 1(b) displays respectively the bulk bands for CNs=0/3/−2/−1 and Fig. 1(c)(d) depict the TESs of zigzag and armchair edge ribbons with CN=3, which are agreement with pioneering work[28]. The near zero-energy $|\psi|^2$ along the zigzag and armchair edge lattice sites manifests Majorana fermions emerging near the lattice boundary as shown in Fig. 1(e).

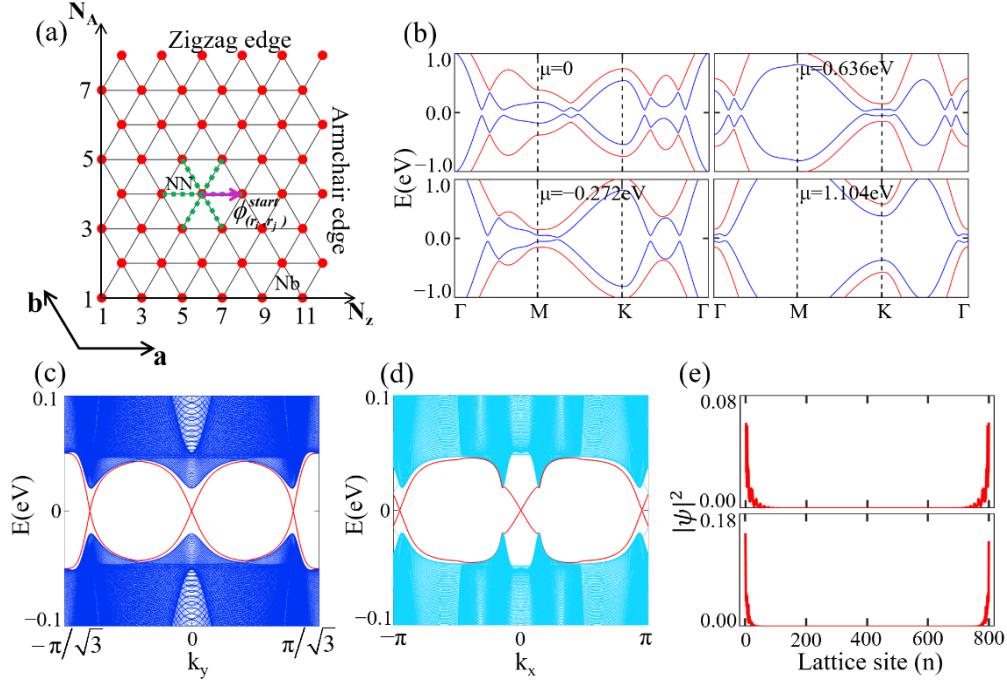

Fig. 1 (a) The trigonal lattice of Nb atoms and zigzag/armchair ribbon with width $N_Z/N_A$. The green dashed lines represent six nearest neighbors of an Nb atom. The pink arrow denotes the initial projection vector. (b) The bands of bulk NbSe$_2$ with $\alpha_R$=0.1, $\Delta_s$=0.05, $V_z$=0.1eV. The CNs are 0($\mu$=0), 3($\mu$=−0.272eV), −2($\mu$=0.636eV), −1($\mu$=1.104eV). (c)(d) Edge spectra for (c) zigzag and (d) armchair ribbons with 800 lattice sites, respectively. The $\Delta_s$=0.05eV, $\alpha_R$=0.1eV, $\mu$=−0.272eV and $V_z$=0.1eV are adopted. (f) The upper and lower panels are the $|\psi|^2$ around the Fermi level (E$_F$) for zigzag (d) and armchair (e) ribbons.

## 3. Results and discussion

To characterize the topological properties of TSCs, we initially calculate the CN as a function of $\mu$ and $V_z$ for the mixed paring state of $\Delta_s$ and $\Delta_{sA1}$, as illustrated in Fig. 2. We study the TSC phase through varying simply the percentage of the two types of pairing potential under the fixed total magnitude of $\Delta_t = \Delta_s + \Delta_{sA1}$ =0.05 or 0.03. Here, we use the p$_s$⊕p$_{sA1}$ to denote the mixed paring potential, where p$_s$ presents thepercentage for s-wave $\Delta_s$ and p$_{sA1}$ represents the percentage for NN paring $\Delta_{sA1}$, with p$_s$+ p$_{sA1}$=1. Previous study shows bulk NbSe$_2$, considering a pure s-wave superconducting pairing and RSOC, exhibits TSC states with non-zero CNs as long as the $V_z$ exceeds the $\Delta_s$ under appropriate $\mu$[28]. For 0.98⊕0.02 mixed superconducting state，monolayer NbSe$_2$ still exists three TSC phases with non-zero CNs 3/−2/−1. However, the three regions of TSC changes in a non-consistent manner due to the mixed effect of s-wave and extend s-wave ($\Delta_{sA1}(\boldsymbol{k}) \approx 2-\frac{3}{2}|\boldsymbol{k}|^2$) pairings, as depicted in Fig. 2(a). When the $\Delta_t$ is 0.05, the regions with CNs= 3 and −2 are larger, while the region with CN= −1 is smaller than the pure s-wave pairing. When the total superconducting pairing decreases ($\Delta_t$ =0.03), the three regions with topological nontrivial CNs the similar change trend. We present the results under total mixed superconducting pairing of 0.05 in the following calculations, unless specified otherwise. Furthermore, our findings do not change qualitatively with pairing strength if total mixed superconducting pairing is not extremely small.

As the proportion of $\Delta_{sA1}$ ($p_{sA1}$) increases, the regions with CNs= 3 and −2 become larger and larger, and the region with CN= −1 becomes progressively smaller. More interestingly, we clearly observe that the critical $V_z$ to generate TSC phases with CNs=3/−2 decreases significantly, indicating a substantialreduction in the experimental difficulty of realizing the TSC phases. When the $p_{sA1}$ increases to 0.14, which is the first critical proportion in our calculations, the region with CN= −1 disappears under $V_z$= [0.0, 0.1]. The mimimal (critical) $V_z$ needed to form TSC phases with CN=3/−2 is 0.017/0.003, which is significantly less than the critical $V_z$ compared with the pure s-wave case as seen in Fig. 2(b). Fig.S1 shows more TSC phases with non-zero CNs (e.g., CN=±6, ±5, ±4, ±3, ±2, ±1) appearing and the TSC phase regions with CNs = 3 and −2 start to shrink when $p_{sA1}$ is greater than 0.14. The CN displays a sensitive dependence on the $V_z$ and μ and distributes irregularly between CN = 3 and CN=−2 phases, which has also been observed in previous studies[15, 17-21, 25, 31]. It is notably that the sensitive non-zero CNs appear even under the $V_z$ =0 when $p_{sA1}$ is greater than or equal to 0.16 (see Fig.S1-S3 and Fig. 2(c)(d)) , suggesting that there may be more intrinsic superconductors (e.g. 1T-PdSe$_2$[32], 1T-PdTe$_2$[33], 2H-TaS$_2$[13], etc.) that can induce TSC phases through neighbor pairing without external tuning approaches(including strain, gating and magnetism). Fig. S1 displays the region with sensitive CNs steadily expands, and the easy-to-vary in CNs become lesser sensitive as the $p_{sA1}$ increases. When $p_{sA1}$ increases to 0.20, the region originally with CN=3 disappears and is replaced by a CN=−3. At this second critical proportion ($p_{sA1}$=0.20), the whole phase diagram mainly includes TSC phases with irregularly nonzero CNs at $V_z$=0 and non-zero CNs=−3, 6, −2. When $p_{sA1}$ is greater than 0.20, the regions with CN =−3 and 6 become sensitive again. Moreover, the size of these regions decreases continually, and the incipient region with CN=3 reappears, as shown in Fig. 2(c) and Fig. S2. As $p_{sA1}$ increases progressively to 1.00, only TSC phase with the nonzero sensitive CNs near μ=0 exists in the phase diagram, and all other TSC phase regions with fixed CNs disappear, as presented in Fig. 2(d) and Fig. S3.

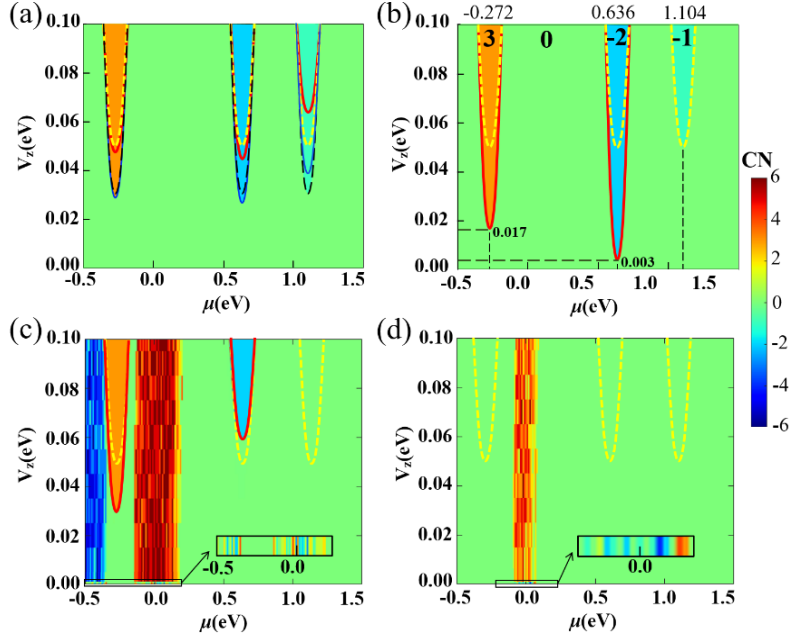

Fig. 2 The TSC phase diagrams of NbSe$_2$ with mixed pairing of $\Delta_s$ and $\Delta_{sA1}$. The yellow (black) dashed lines represent the TSC phases with pure s-wave paring $\Delta_s$=0.05 (0.03) and $\Delta_{sA1}$=0. The red (blue) solid lines represent the TSC phases with total mixed paring $\Delta_t$=0.05 (0.03). (a) 0.98⊕0.02, (b) 0.86⊕0.14, (c) 0.70⊕0.30, (d) 0.00⊕1.00.

To verify the topological nature of the TSCs, we construct a tight-binding model for an infinitely long strip of NbSe$_2$ with finite width 800 along zigzag and armchair edges (Fig. 1(a)). Fig. 3 and Fig. 4 show the energy spectra of zigzag and armchair edge ribbons under mixed superconducting states, respectively. Our calculations demonstrate mismatch between sensitive CNs and corresponding TESs, as shown in Fig. 3(c)(d)(i)(j) and Fig. 4(c)(d)(i)(j). As the p$_{sA1}$ increases, the correspondence of CNs and TESs of the studied system undergoes three stages: from matching to mismatching, then to mix matching with mismatching. Here, we mainly focus on the TSC phases with fixed CN=−2 and 3 because its corresponding μ~0.636 and −0.272eV is near the E$_F$ (Table I).

In the first stage, the whole phase diagram consists of the TSC phases with CNs=0, −1, −2, 3. The pairing percentage varies from 1.00⊕0.00 to 0.84⊕0.16 for TSC phase with CN=3 and from 1.00⊕0.00 to 0.88⊕0.12 TSC phase with CN=−2, which indicates TESs in CN=−2 TSC phase are more susceptible than in CN= 3. The CNs are equal to the number of TESs on zigzag and armchair edge ribbon as shown in Fig. 3(a)(b) and Fig. 4(a)(b), which is agreement with the normal BBC[2, 34].

In the second stage, the whole phase diagram contains not only fixed CNs (=−1, −2, 3), but also sensitive CNs (=±6, ±5, ±4, ±3, ±2, ±1) regions (see Fig. S1 and Fig. S2). The mixed pairing percentage varies from 0.83⊕0.17 to 0.74⊕0.26 for CN=3 TSC phases and from 0.87⊕0.13 to 0.82⊕0.18 for CN=−2 TSC phases. In this stage, the BBC of all TSC phases with topological nonzero CNs is broken (Fig. 3(c)(d), Fig. 4(c)(d) and Fig. S4). The mismatch between CN and TESs may results from the hybrid effect of bulk and edge states due to mini-bulk-band-gap, which will be discussed in detail later. In third stage, fixed CN regions re-match the BBC (Fig. 3(e)(f)(h)(g) and Fig. 4(e)(f)(h)(g), while easy-to-vary CN regions manifest a breaking of the BBC (Fig. S5).

Compared with the first stage aforementioned, in the third stage, the phase diagram includes simultaneously robust- and fragile-TESs, both consistent with the BBC as shown in Fig. 3(e)(f)/Fig. 4(e)(f) and Fig. 3(h)(g)/Fig. 4(h)(g), respectively. For robust TESs, the edge bands connecting conduction bands and valence bands are nontrivial and do not eliminate due to the shifting of the $E_F$ (Fig. 3(e)(f) and Fig. 4(e)(f)). For fragile-TESs, the bands connect the conduction/valence bands themselves or form isolated in-gap bands, therefore can be easily removed by changing the $E_F$ or a continuous perturbation as depicted in Fig. 3(h)(g) and Fig. 4(h)(g). Noteworthy, at $V_z=0$, the TESs for 0.00⊕1.00 state (see Fig. 3(i)(j) and Fig. 4(i)(j)) exhibit robustness, which verifies the above TSC phase diagrams and facilitates experimental detection and the development of TESs-based topological quantum computation[7].

Table I The correspondence between CNs (=3, −2) and TESs under different percentage $p_s$ and $p_{sA1}$.

| TSC phases with CN=3 | | | TSC phases with CN=−2 | | |
| --- | --- | --- | --- | --- | --- |
| $p_s$ | $p_{sA1}$ | correspondence | $p_s$ | $p_{sA1}$ | correspondence |
| 1.00~0.84 | 0.00~0.16 | matching | 1.00~0.88 | 0.00~0.12 | matching |
| 0.83~0.74 | 0.17~0.26 | mismatching | 0.87~0.82 | 0.13~0.18 | mismatching |
| 0.73~0.42 | 0.27~0.58 | matching | 0.81~0.60 | 0.19~0.40 | matching |

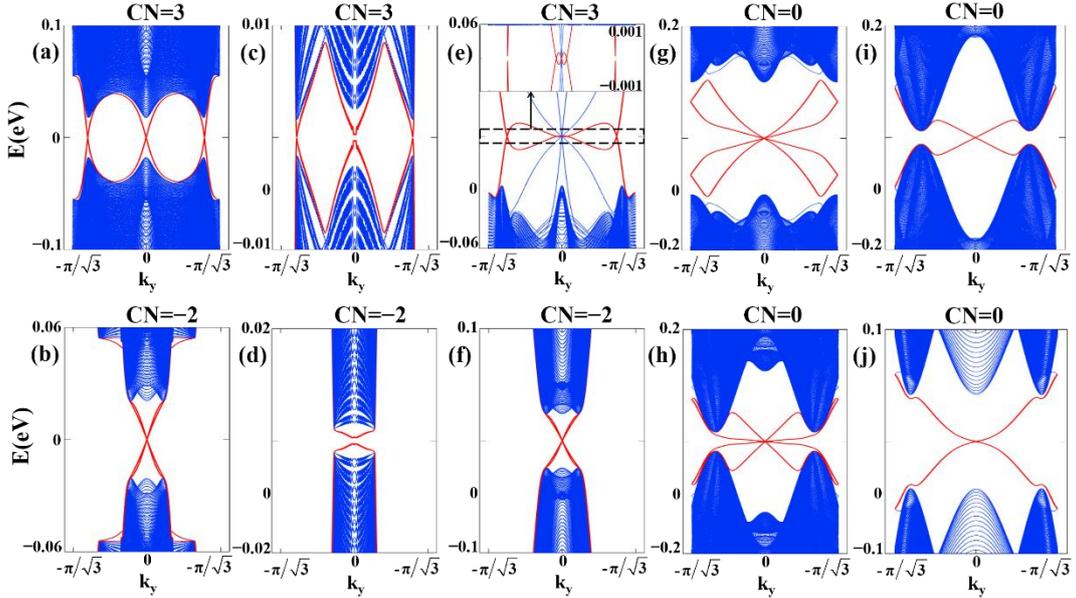

Fig. 3 The TESs of zigzag ribbon with 800 unit cells under various mixed pairing of $\Delta_s$ and $\Delta_{sA1}$, $\alpha_R$=0.1eV, (a)-(h)$V_z$=0.1eV and (i)-(j)$V_z$=0.0eV. (a)(c)(e)(d)μ=−0.272eV, (b)(d)(f)μ=0.636eV, (h) μ=−0.1eV, (i)μ=−0.5eV and (j)μ=−0.11eV. (a)(b)0.98⊕0.02, (c)0.82⊕0.18, (d)0.86⊕0.14, (e)0.52⊕0.48, (f)0.70⊕0.30, (g)-(j)0.00⊕1.00.

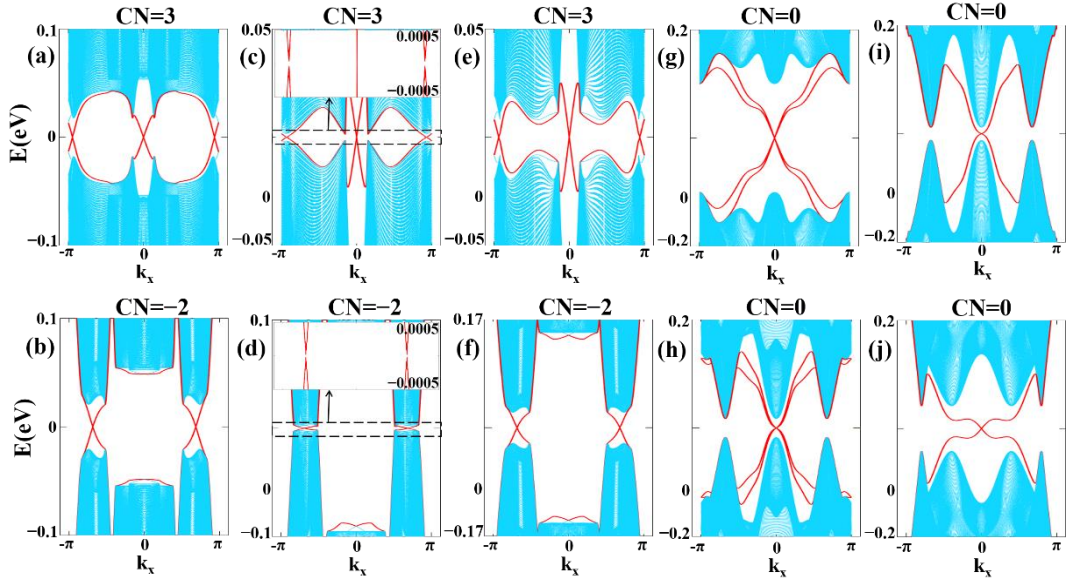

Fig. 4 The TESs of armchair ribbon with 800 unit cells under various mixed pairing of $\Delta_s$ and $\Delta_{sA1}$, $\alpha_R$=0.1eV, (a)-(h)$V_z$=0.1eV and (i)-(j)$V_z$=0.0eV. (a)(c)(e)(d)$\mu$=−0.272eV, (b)(d)(f)$\mu$=0.636eV, (h) $\mu$=−0.1eV, (i)$\mu$=−0.5eV and (j)$\mu$=−0.11eV. (a)(b)0.98⊕0.02, (c)0.82⊕0.18, (d)0.86⊕0.14, (e)0.52⊕0.48, (f)0.70⊕0.30, (g)-(j)0.00⊕1.00.

The proximity effect between topological insulators and superconductors induces topological superconducting phases[35-40]. However, there are non-zero energy edge states inherited from the topological insulator in TSCs, affecting CNs and resulting in the mismatch[23]. Therefore, CN can predict the number of TESs crossing the $E_F$ but cannot predict the number of zero-energy Majorana edge states[17]. The closure and opening of bulk band-gap can lead to a topological phase transition[21], while the narrowing and widening of the bulk band-gap can also affect whether the TESs remain clean. We calculate the band-gap between conduction and valance bands to illustrate the effect of the mixed superconducting pairing function on TESs as shown in Fig. 5(a)(b)(c). The $|\psi|^2$ can exhibit topological superconductivity in real space near the $E_F$, as depicted in Fig. 5(d)-(i). The bulk band-gap of TSC phases with CNs=3 and −2 oscillates with increasing $p_{sA1}$, leading to the three stages mentioned above between CN and TESs (Fig. 5(a) and Table I).

In the first stage with $p_{sA1}$~[0.00, 0.16]/[0.00,0.12], the bulk band-gaps of TSC phases with CNs=3/−2 are larger than ~0.01eV orders of magnitude and lead to weak hybrid interaction between the bulk projection states and TESs, consequently demonstrating a good BBC.(see Fig. 3(a)(e) and Fig. 4(a)(e)). The $|\psi|^2$ is localized dominantly in the both edges of ribbon, as seen in Fig. 5(d) and Fig. S6(a)(e)(h). At the second mismatching stage, the bulk band-gaps are tiny and remain on the order of $10^{-4}$eV within the range $p_{sA1}$=[0.17, 0.26] for CNs=3 and [0.13,0.18] for CN=−2 TSC phases, respectively (Fig. 5(a)(c)). There are many non-topological bulk bands near the $E_F$ due to the tiny band-gap, resulting in a large amount of edge projection states. Then the projection states and TESs hybrid strongly and cause the delocalization of TESs, i.e., the large $|\psi|^2$ on the whole ribbon (Fig. 5(e) and Fig. S6(b)(f)(i)), consequently the breaking of BBC. The $|\psi|^2$ shows an oscillating behavior with relatively high-amplitude from the edge to the center of ribbon (see Fig. 5(e) and Fig. S6(b)(f)(i)) compared with the localization feature of TESs in the first stage (see Fig. 5(d) and Fig. S6(a)(e)(h)). The oscillating behavior is generally originated from the interference between bulk states and TESs[41]. However, the bulk band-gap re-increases and reaches its maximum at $p_{sA1}=0.40$ for CN=3 and $p_{sA1}=0.34$ for CN=−2 TSC phases, with the increasing $p_{sA1}$ at the third matching stage. It is accompanied by a gradual reduction in non-topological bands near the $E_F$ and TESs re-emerge and the $|\psi|^2$ re-localizes (see Fig. 5(f)(g)(h)(i) and Fig. S6(c)(g)(j)), although the band-gap varies.

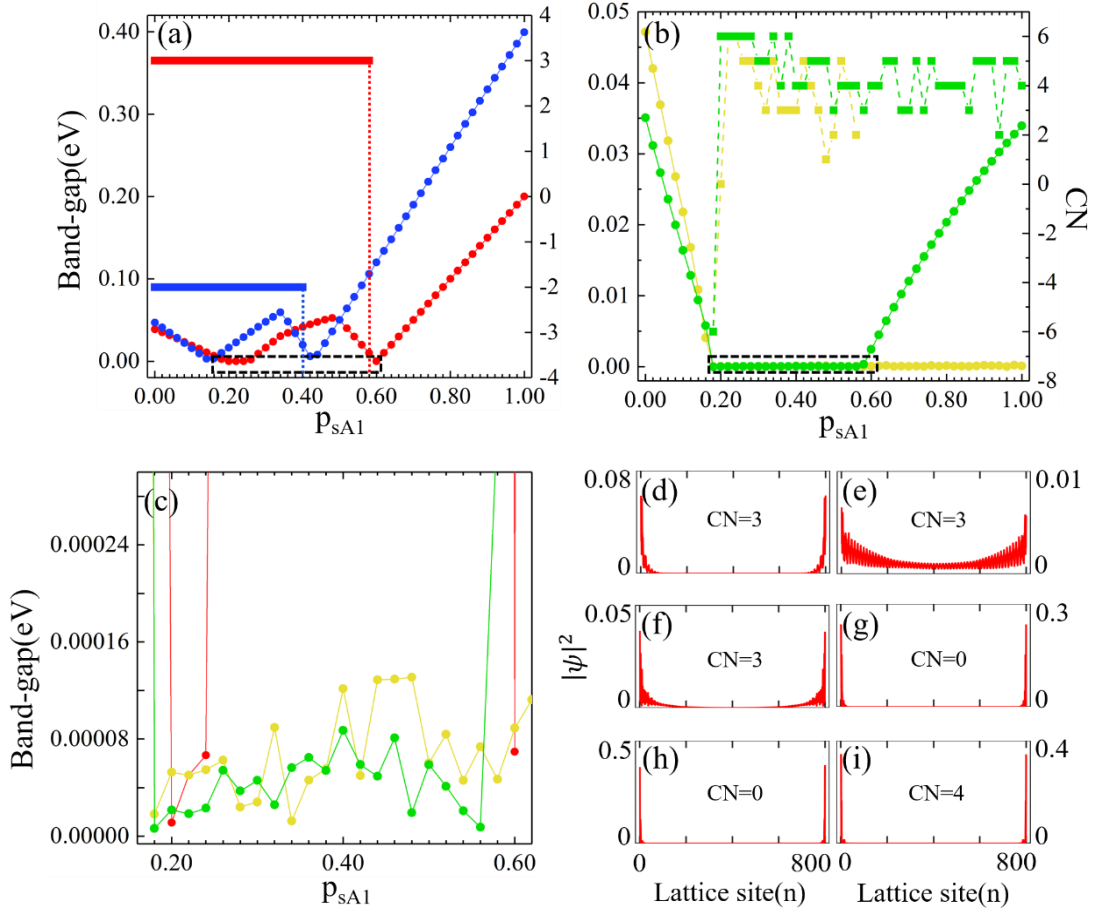

Fig.5 (a)-(b) The bulk band-gap and CN of different TSCs under $\alpha_R=0.1$eV and $V_z=0.1$eV. Circles represent the bulk band-gap and squares represent the CNs. Red, blue, green and yellow mark μ=−0.272eV, 0.636eV, −0.1eV, 0.0eV, respectively. (c) The enlarged bulk band-gap in the black dashed line boxes of (a)(b). (d)-(i) are the $|\psi|^2$ of zigzag ribbon near the $E_F$ in real space. (d)-(g) μ=−0.272eV (d)0.98⊕0.02, (e)0.82⊕0.18, (f)0.70⊕0.30 and (g)0.00⊕1.00. (h)-(i)0.00⊕1.00 (h)μ=−0.10eV and (i)μ=0.00eV.

## 4. Conclusions

In conclusion, we systematically investigate TSCs in monolayer NbSe$_2$ by considering various proportions of mixed pairing $\Delta_s$ and $\Delta_{sA1}$ based on a tight-binding model. Firstly, for mixed superconducting pairing states, we observe rich phases with fixed and susceptible CNs as the variation of the chemical potential μ and and out-of-magnetic field $V_z$. The TSC phase manifests matching and mismatching features in relation to the existence of a BBC as the $p_{sA1}$ increasing. Secondly, the results clearly demonstrate a significant reduction in the critical $V_z$ required to induce TSC phases due to the mixed effect of $\Delta_s$ and $\Delta_{sA1}$. Intriguingly, the TSC phases can be modulated at $V_z=0$ under appropriate μ and $p_{sA1}$ (when $p_{sA1}$ is greater than or equal to 0.16). Moreover, the reality of TSC phase at $V_z=0$ is further verified by calculating the robust TESs along zigzag and armchair ribbons. Thirdly, it is demonstrated that the CNs do not always match with the number of TESs resulting from the hybrid effect of bulk state and edge state, as evidenced by calculating the bulk band-gap as a function of the $p_{sA1}$. It becomes evident that bulk state hybridize with edge state when the band-gap is tiny, leading to a mismatch between CNs and TESs. Finally, by calculating the $|\psi|^2$ along ribbon site, it is found that TESs are localized at the boundary for a matching BBC. However, there is an oscillating behavior of slow decay with the lattice sites for the mismaching/broken BBC. Our investigation provides a new idea and an easier way for the design and regulation of TSC materials in experiment, as well as theoretical guidance for the fabrication of TSC quantum devices.

**Acknowledgements**

This work is supported by the National Natural Science Foundation of China (Grant No. 11204261), a Key Project of the Education Department of Hunan Province (Grant No. 19A471), Natural Science Foundation of Hunan Province (Grant No. 2018JJ2381).